\documentclass[epjST]{svjour}
\pdfoutput=1
\usepackage{geometry}
\usepackage[T1]{fontenc}
\usepackage{lmodern}

\usepackage[utf8]{inputenc}
\usepackage{amsmath, amssymb}
\usepackage{physics}
\usepackage{graphicx}
\usepackage{caption}
\usepackage{subcaption}
\usepackage{float}
\usepackage{todonotes}
\usepackage[comma,numbers,sort&compress]{natbib}
\usepackage{hyperref}
\usepackage{booktabs}
\usepackage[rightcaption]{sidecap}
\sidecaptionvpos{figure}{t}

\usepackage{siunitx}
\usepackage{xspace}

\graphicspath{{./figures/}}

\newcommand{\Hone}{$\mathcal{H}_0^1$\xspace}
\newcommand{\Htwo}{$\mathcal{H}_0^2$\xspace}
\newcommand{\Hthree}{$\mathcal{H}_0^3$\xspace}
\newcommand{\Hfour}{$\mathcal{H}_0^4$\xspace}

\newcommand{\diff}{\mathop{}\!\mathrm{d}}

\begin{document}
\title{Detecting contagious spreading of urban innovations on the global city network}

\author{Niklas H. Kitzmann\inst{1,2}\fnmsep\thanks{\email{kitzmann@pik-potsdam.de}, ORCID: 0000-0003-4234-6336}
\and Pawel Romanczuk\inst{3,4}\fnmsep\thanks{ORCID: 0000-0002-4733-998X}
\and Jonathan F. Donges\inst{1,5}\fnmsep\thanks{ORCID: 0000-0001-5233-7703} 
}
\institute{Earth System Analysis, Potsdam Institute for Climate Impact Research, Member of the Leibniz Association, 14473 Potsdam, Germany, EU \and Institute for Physics and Astronomy, University of Potsdam, 14476 Potsdam, Germany, EU \and Institute for Theoretical Biology, Humboldt University of Berlin, 10115 Berlin, Germany, EU \and Bernstein Center for Computational Neuroscience, Humboldt University of Berlin, 10115 Berlin, Germany, EU \and Stockholm Resilience Centre, Stockholm University, 10691 Stockholm, Sweden, EU}

\abstract{
Only a fast and global transformation towards decarbonization and sustainability can keep the Earth in a civilization-friendly state. 
As hotspots for (green) innovation and experimentation, cities could play an important role in this transition.
They are also known to profit from each other's ideas, with policy and technology innovations spreading to other cities.
In this way, cities can be conceptualized as nodes in a globe-spanning learning network. 
The dynamics of this process are important for society's response to climate change and other challenges, but remain poorly understood on a macroscopic level.
In this contribution, we develop an approach to identify whether network-based complex contagion effects are a feature of sustainability policy adoption by cities, based on dose-response contagion and surrogate data models. 
We apply this methodology to an example data set, comprising empirical data on the spreading of a public transport innovation (Bus Rapid Transit Systems) and a global inter-city connection network based on scheduled flight routes.
We find evidence pointing towards a contagious spreading process which cannot be explained by either the network structure or the increase in global adoption rate alone.
This suggests that the actions of a city's abstract "global neighborhood" within the network of cities may be an important factor in which policies and innovations are implemented, with potential connections to the emergence of social tipping processes.
The methodology is generic, and can be used to compare the predictive power for innovation spreading of different kinds of inter-city network connections, e.g. via transport links, trade, or co-membership in political networks.
} 
\maketitle
\newpage
\section{Introduction}%
\label{sec:intro}
Anthropogenic climate change and other human-made pressures on the environment are threatening the civilization-friendly Holocene state that the Earth system has been existing in for thousands of years \cite{rockstrom_planetary_2009,steffen_planetary_2015}.
A world-wide transformation of the economic, societal and political world towards sustainable practices and technologies is urgently needed to mitigate these effects \cite{otto_social_2020}.
Cities may play an important role as active agents in this global sustainability transformation, with their potential for impactful change mediated by a number of factors:
Precisely because some of the largest drivers of planetary boundary transgressions are located in cities, their mitigation potential is high.
For example, the urban population accounts for over \SI{80}{\percent} of global greenhouse gas emissions \cite{hoornweg_cities_2011}, and has large freshwater \cite{paterson_water_2015} and chemical pollution  footprints \cite{brooks_toxicology_2020}.
Already, over half of the global population lives in cities, a share that is projected to rise to two thirds by the middle of the 21st century \cite{united_nations_world_2019}.
These urban populations are also among the most threatened by negative consequences of environmental changes, such as flooding events \cite{skougaard_kaspersen_comparison_2017} and heat waves \cite{li_synergistic_2013}.
Thus, if the challenge of remaining within the planetary boundaries can be solved, it must be solved in the urban context.
At the same time, cities are uniquely suited to address these issues.
As centers of knowledge and innovation \cite{knight_knowledge-based_1995,bettencourt_growth_2007}, and possessing significant economic resources \cite{taylor_specification_2001}, innovative solutions are most likely to originate here.
This is especially true for sustainability innovations, where so-called frontrunner cities have shown how experimentation with sustainability can create positive inertia for change \cite{sengers_experimenting_2019, irvine_positive_2019, castan_broto_survey_2013}.

To become relevant for the global sustainability transformation, locally-conceived and implemented innovations must spread world-wide.
As part of the human response to anthropogenic environmental changes, 
the spreading of sustainability innovations thus represents an important Earth system \cite{reid_earth_2010} process.
Innovation and policy transfer has been extensively studied in the social sciences  \cite{benson_what_2011,dolowitz_who_1996, dolowitz_learning_2000, marsden_how_2011}, primarily in case studies investigating individual cities or small city networks \cite{marsden_transfer_2010,  pojani_going_2015, lee_who_2012}.
While this focus on individual circumstances is surely well-placed for studying such complex and diverse processes, we argue that it would be well complemented by data- and model-based studies of the macroscopic global dynamics of urban innovation spreading.

In this contribution, we develop a method for investigating the proliferation of urban sustainability innovations on a global scale.
Drawing on methods from (social) network science, we conceptualize the innovation transfer as a spreading process on the global network of cities.
Specifically, we hypothesize a complex contagion process \cite{dodds_generalized_2005,centola_complex_2007}, implying a non-trivial relationship between the probability of a city to implement an innovation and its exposure(s) to it from other cities.
Networked spreading and contagion processes have been studied in many different scientific subjects, such as epidemics \cite{ daley_epidemic_1999}, cascading failures  \cite{buldyrev_catastrophic_2010}, and the formation and spreading of social norms, opinions and behaviors \cite{nyborg_social_2016,tsvetkova_social_2014,rosenthal_revealing_2015}.
The spreading of social, political and technological innovations relevant for sustainability transitions and rapid decarbonisation have also been identified as a promising approach for understanding the emergence of social tipping points in this context \cite{tabara_positive_2018, farmer_sensitive_2019,otto_social_2020,sharpe_upward-scaling_2021}.

In line with the literature, throughout this work we use terms such as ``infection'' and ``contagion'' to describe the innovation spreading process.
In this context, the ``infection'' of a city should simply be understood as the adoption of the studied innovation by a city, without negative connotations or the inference of passivity.
Likewise, ``contagion'' only implies that this process may be directly influenced by other cities which have already implemented the innovation, and which are connected to the original city in some way.

We use a set of example data sets to develop and demonstrate our method.
As a proxy for the inter-city links facilitating the spreading of innovations, we use the global network of scheduled flight routes.
We correlate this (static) network with the spreading of Bus Rapid Transit Systems (BRTs), a public transport innovation which combines features of bus networks and light rail systems \cite{levinson_bus_2002}.
The spreading of BRTs has previously been investigated in case studies \cite{wood_politics_2015, matsumoto_analysis_2007}, but never on a global scale.

Detecting contagion in such low-rate spreading processes on a large, high-density network poses a statistical challenge.
Because of the small number of total infections, we cannot rely on interpreting the functional shape of the infection rate as in \cite{rosati_modelling_2021}.
Instead, our approach is based on dose response functions (DRFs) \cite{haas_conditional_2002, dodds_universal_2004, dodds_generalized_2005}, an analytical tool that has been used to study a variety of simple and complex spreading processes, such as the diffusion of information on social media networks \cite{hodas_simple_2014} and the spreading of health-related behaviors among students \cite{donges_dose-response_2021}.
Dose response functions encode the probability of infection of a node, as a function of the exposure (or ``dose'') received from connected nodes.
We modify this measure for a fractional contagion paradigm \cite{rosenthal_revealing_2015}, to describe a city's probability of adopting a new innovation dependent on the fraction of its network neighborhood that has already implemented the innovation.
We then develop a hierarchy of surrogate models, successively excluding non-contagion-related mechanisms that may confound the observation of contagion processes.
The surrogate model method relies on Monte-Carlo-based, data-derived hypothesis tests to analyze specific data features without prescribing concrete underlying mechanisms.
Surrogate models have successfully been used as a tool in exploratory data analyses \cite{vicente_transfer_2011,casdagli_chaos_1992}, particularly for investigating networked processes  \cite{gauvin_randomized_2020, holme_temporal_2012}, including epidemic and social contagion \cite{genois_compensating_2015, karimi_threshold_2013, donges_doseresponse_2021}, and time series data \cite{theiler_testing_1992, schreiber_surrogate_2000}.
Partially randomizing the empirical data in line with specific null hypotheses, and then comparing key measurements (here, DRF functional shapes) allows us to investigate correlations found in the data, and their causal relevance.

This paper is structured as follows: 
In Sect.\,\ref{sec:data}, a description of the data sets used to demonstrate the method is given, along with a motivation for their selection.
This is followed by the description of the employed methods in Sect.\,\ref{sec:methods}, detailing our use of DRFs (Sect.\,\ref{sec:methods_DRF}) and surrogate models (Sect.\,\ref{sec:methods_surrogatedata}).
The results of our analysis are reported in Sect.\,\ref{sec:results}; we discuss them and conclude in Sect.\,\ref{sec:discussion}.

\section{Data}
\label{sec:data}
To demonstrate the methodology, we use one example data set each for the city network and the spreading innovation, respectively.
The choice for these illustrative data sets is driven by three considerations: the availability of an adequate data set, the plausibility of finding contagious spreading behavior in this data, and the scientific or social relevance of understanding the spreading process of the system itself and its analogues.
The data sets used here are the global network of scheduled flight routes for the network component, and the adoption of Bus Rapid Transit System (BRT) public transport innovation for the spreading component.

\subsection{Flight Route Network}
The choice of flight route connections for the city network component has several advantages.
First, an globally homogeneous data set is available from public sources \cite{openflights}.
This data also includes smaller cities, which are often not the focus of city network research \cite{taylor_specification_2010}.
Furthermore, we expect any kind of city-to-city connection, be it on an economic, political, or cultural level, to also produce some amount of flight traffic.
The flight route network can thus serve as a plausible proxy for any underlying inter-city linkages.

\begin{figure}[h]
    \centering
    \includegraphics[width=0.495\linewidth]{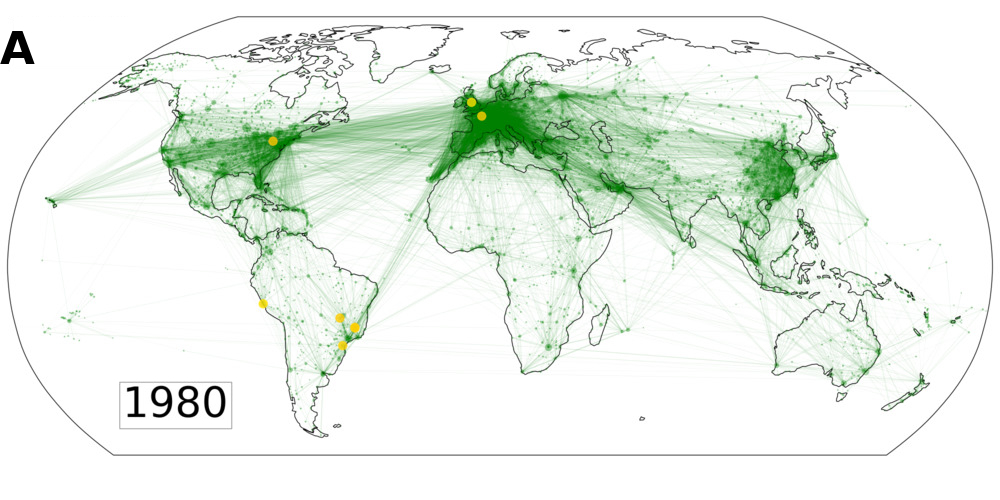}\hfill
    \includegraphics[width=0.495\linewidth]{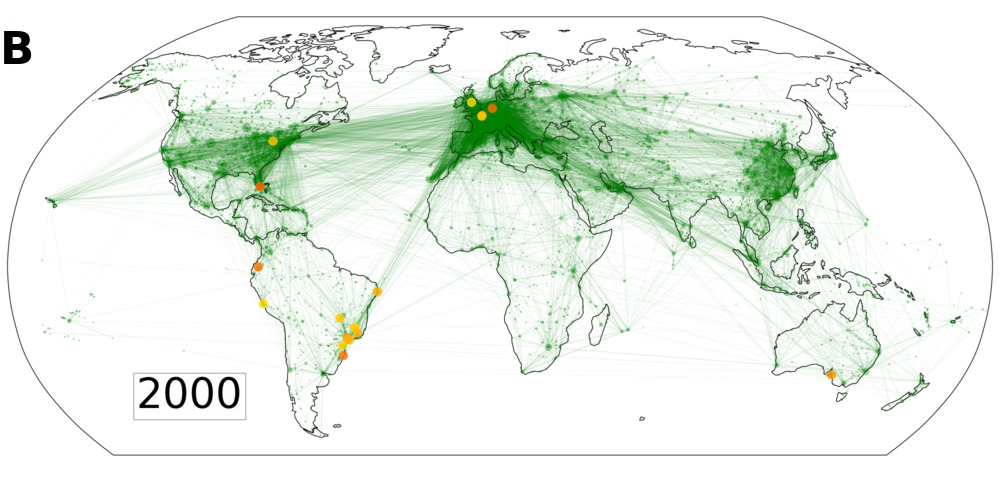}
    \raisebox{0.15cm}{\includegraphics[width=0.495\linewidth]{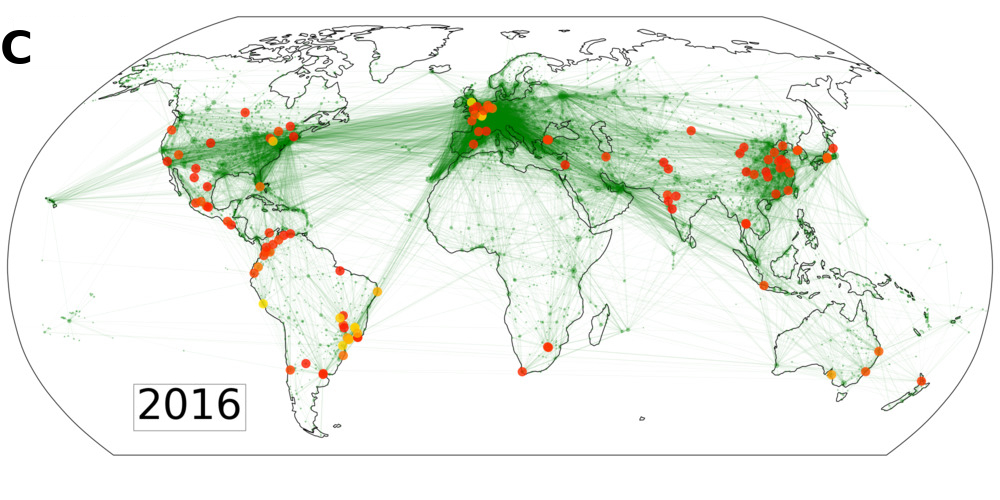}}\hfill
    \includegraphics[width=0.42\linewidth]{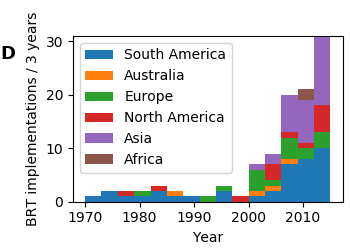}\hspace{0.8cm}
    \caption{\textbf{A-C} Visualization of the spreading of Bus Rapid Transit Systems in 1980 (\textbf{A}), 2000 (\textbf{B}) and 2016 (\textbf{C}); the latter two represent the bounds of the time interval investigated here. The date of implementation is displayed on a color scale from yellow (1972) to red (2016). Overlaid in green is the global network of flight routes. \textbf{D} Implementation rate of Bus rapid transit systems in a stacked histogram, color coded by continent. A marked rise of implementations is apparent after the year 2000, prompting our scrutiny of this time interval.}
    \label{fig:BRTspreading}
\end{figure}

We source the network data from a publicly available data set \cite{openflights}, which contains information on airports and scheduled routes that are visualized in Fig.\,\ref{fig:BRTspreading}. 
We correlate this information with data on city locations and population sizes \cite{wikidata}, to transform the airport-to-airport direct route network into an undirected, weighted city-to-city network.
The exact algorithm for calculating a city's connection strength to another is described in App. \ref{app:city_distances}.
Only cities with a population of greater than \num{60000} are considered here, corresponding to the lowest population threshold which includes almost all cities which have implemented a Bus Rapid Transit System (see Sect. \ref{sec:data_brt}).
As the flight route data source provides a snapshot of the global flight route network dated to 2014, we assume the network to be static. 
Limitations imposed by this choice are discussed in Sect.\,\ref{sec:discussion}.

\subsection{Bus Rapid Transit Systems}
\label{sec:data_brt}
We choose the implementation of Bus Rapid Transit Systems (BRT) as the spreading innovation component of the illustrative analysis.
BRTs are a public transport innovation, first developed in the early 1970's \cite{brtdata}.
Combining a number of measures such as dedicated bus lanes, frequent service with timed transfers, off-board fare collection, and preferential intersection treatment, BRTs are frequently compared to light rail networks \cite{levinson_bus_2002}.
Often representing a cost-effective way of implementing a high-quality public transport network for cities, they can play an important role in shifting the modal share towards environmentally friendlier means of transport \cite{deng_recent_2011}.

A comprehensive database of BRT implementations is jointly maintained and publicly provided by the BRT+ Centre of Excellence and EMBARQ, the WRI Ross Center for Sustainable Cities signature initiative for sustainable transport \cite{brtdata}.
Only implementations rated ``Bronze'', ``Silver'' or ``Gold'' by these organizations are considered in this analysis, in order to exclude systems that only share a limited amount of features with full BRT implementations.
The global implementation rate of BRTs is displayed in Fig.\,\ref{fig:BRTspreading}.
Following several decades of low adoption rates, a marked increase can be observed after the year 2000. 
To better understand this phenomenon, and exclude times of low activity that may drown out any potential contagion effects in the data, we focus on this ``epidemic" phase of rapidly rising implementations between the years 2000 and 2016.
At the beginning of this period, BRTs are already present on four continents (Fig.\,\ref{fig:BRTspreading}B).
The data is considered in a time-stepped fashion, with time step length $t=1$ year.

\section{Methods}
\label{sec:methods}
We use a dose-response-contagion approach to investigate contagion effects, described in Sect.\,\ref{sec:methods_DRF}
To differentiate between true contagion effects and confounding factors like homophily and shared environments, we use a hierarchy of surrogate data sets.
This allows for a search for evidence of causal contagion effects by excluding alternative hypotheses, and is described in Sect.\,\ref{sec:methods_surrogatedata}.

\subsection{Dose-Response Functions}
\label{sec:methods_DRF} \label{sec:errorbars}
Dose-response functions (DRFs) are a useful tool in characterizing contagion effects on networks \cite{haas_conditional_2002, dodds_universal_2004, donges_doseresponse_2021}.
They represent the functional dependence of a node's infection likelihood $p_\text{inf}$ on the exposure ``dose'' $I$ from neighboring infected nodes.
Depending on the underlying contagion process, DRFs can have different functional forms, such as smooth, sigmoidal curves, and even sharp step-like functions for threshold-based contagion processes \cite{dodds_universal_2004}.
As a measure of the ``dose'' received by a city, we define the infection pressure $I_i(t)$ experienced by a city $i$,
\begin{equation}
    \label{eq:infection_pressure}
    I_i(t)=\sum\limits_{j}^{N_i} w_{i,j}s_j(t)\left(\sum\limits_{k}^{N_i}w_{i,k}\right)^{-1}\, .
\end{equation}
Here, $N_i$ represents the number of cities connected to city $i$, and $w_{i,j}$ holds the weight of the connection between cities $i$ and $j$. 
The infection status $s_j(t)$ is 1 or 0 if city $j$ is infected or uninfected at time $t$, respectively.
This definition represents a fractional contagion paradigm, a type of complex contagion \cite{rosenthal_revealing_2015} that features the inhibition of infection probability by non-infected neighbors.
It is motivated by the conjecture that cities with a high network degree should be less likely to adopt the innovation after an exposure of a certain strength, than cities that have few or low-weight other connections.
In this way, high numbers of connections to non-infected cities may ``drown out'' the effects of infected neighbors.
Vice versa, exposures are more effective if they are experienced by a city with a low degree of connectivity.
This also solves the problem of high-degree nodes, such as the air traffic hub London, which would otherwise receive very high exposures, but cannot plausibly be expected to become infected much more readily.
The system thus retains the intuitive assumption that, for an illustrative example of two connected cities with strongly differing connectivities, an infection is more likely to jump from the high-degree city to the low degree city than vice versa.
This is despite the weight of the connection $w_\text{A,B}=w_\text{B,A}$ remaining symmetric.

From all cities' time series $I_i(t)$ and network connections $w_{i,j}$, we compute the total distribution of infection doses $N(I)$.
Likewise, we compute the distribution of ``successful'' infection doses $C(I)$, comprising only the infection doses $I(\hat t)$ received by cities that became infected at $\hat{t}+1$.
With these two distributions, the node probability of infection $p_\text{inf}$ in response to an experienced infection pressure $I$ is estimated as 
\begin{equation}
    \label{eq:DRF}
    p_\text{inf} (I) \approx \frac{C(I)}{N(I)}\,.
\end{equation}

The DRF $p_\text{inf}$ is an estimator of the true probability of infection after exposure to the infection pressure $I$.
We assume the individual exposure responses to be statistically independent, and can thus understand this process as a series of $N(I)$ independent Bernoulli-experiments.
As we expect the success rate to be very low in our exemplary data set, we estimate the confidence intervals using the Agresti-Coull method \cite{agresti_approximate_1998}, ensuring that the two-sided confidence bounds remain within the $(0,1)$ interval.

\subsection{Surrogate Models}
\label{sec:methods_surrogatedata}

In the presence of confounding effects such as homophily and shared local or global influences, contagion can be hard to identify \cite{shalizi_homophily_2011}.
In the next section (Sect.\,\ref{sec:surrogate_method}), we describe our use of surrogate models to address this challenge, followed by the description of the surrogate models produced for this study (Sect.\,\ref{sec:surrogate_production}).

\subsubsection{Surrogate model method}
\label{sec:surrogate_method}
The surrogate model approach is a statistical proof-by-contradiction method used for investigating specific features and correlations in empirical data sets.
It is based on testing composite null hypotheses on data sets that are derived from the empirical data using Monte Carlo methods \cite{theiler_testing_1992, schreiber_surrogate_2000, gauvin_randomized_2020, holme_temporal_2012}.
A variety of time series \cite{donges_doseresponse_2021, venema_statistical_2006, scheinkman_nonlinear_1989, pritchard_dimensional_1995} and network data sets \cite{donges_doseresponse_2021, wiedermann_spatial_2016, maslov_detection_2004, maslov_specificity_2002} have been analyzed using surrogate models.
The method is described in the following paragraph.

First, a composite null hypotheses $\mathcal{H}_0$ is constructed, which specifies a class of processes that may be sufficient to reproduce the observed empirical data.
Ideally, $\mathcal{H}_0$ excludes certain features or correlations in the data, e.g. relating to hypothetical underlying contagion processes.
Based on $\mathcal{H}_0$ and the empirical data, a surrogate data set is then constructed that resembles the original data, but lacks the hypothetical features excluded by $\mathcal{H}_0$.
Partially randomizing the empirical data set, in a way that is consistent with the null hypothesis, is one option of generating such data sets.
This method, referred to as constrained realizations \cite{theiler_constrained-realization_1996}, forces the resulting data set to resemble the empirical data in key statistical measures as directed by $\mathcal{H}_0$.
Specific correlations and data features may thus be selectively removed, without committing to a specific model.
An ensemble of surrogate data sets is produced for each null hypothesis to reduce statistical uncertainties. 
Finally, a discriminating statistic is computed on both the empirical data and the ensemble of surrogate models.
If the empirical value differs significantly from the ensemble of surrogate values, the null hypothesis is rejected.
This can be regarded as evidence that the preserved features are not sufficient to explain the observations, pointing to a more complex underlying mechanism.
By carefully choosing progressively more complex null hypotheses, the nature of this underlying process can be investigated.

Using the dose-response contagion approach, we would like to compare the empirical DRF with those computed on each surrogate model realization. 
We must therefore find a measure that may quantify the differences of a large number of functional shapes.
Comparing with the simple bin-wise average of the surrogate DRFs is not sufficient here, as individual data points within a single surrogate model realization are not statistically independent from each other:
Since the total number of newly infected cities is constant across all realizations, a raised data point in one bin of the DRF will always result in a different one being lowered. 
To instead compare the individual surrogate model DRF's functional forms with the empirical DRF, we perform weighted least-squares fits to each individual surrogate sample's DRF, and compare the resulting fit parameter distribution.

A number of different DRF shapes are plausible \cite{dodds_universal_2004}, and thus the fitted DRF shape in general has to be chosen with care.
However, in the studied data set, only $\mathcal{O}(100)$ cities have adopted the BRT innovation.
Infection doses in the network are thus generally low, and any saturation effects are unlikely to have significant effects.
We therefore expect the empirically determined DRF to be close to linear, and perform the fit using a polynomial of degree one:
\begin{equation}
    \label{eq:fit_function}
    p(I) \approx m\cdot I + b\,.
\end{equation}
In the fits, each bin's data point is weighted with the total number of times the corresponding infection pressure range was observed in data, displayed as the bin height in Fig.\,\ref{fig:emp_DRF}A.
The fit parameters  $m$ (DRF slope, a measure of the sensitivity of cities' reactions to the dose $I$) and $b$ (y-axis intersection point, a measure of spontaneous background infection rate at received dose $I=0$) are chosen as the discriminating statistic for comparing empirical and surrogate model DRFs.

The fit parameter distribution of the DRFs computed on surrogate models can be visualized as a two-dimensional histogram.
We non-parametrically estimate the underlying two-dimensional probability distribution $P(m,b)$ using kernel density estimation, with Gaussian kernels whose width is set by Silverman's rule \cite{silverman_density_1986}.
We then calculate the value of the quantile function $Q(m_\text{emp}, b_\text{emp})$ of this distribution, integrating the probability density function over the parameter space where it gives a lower probability than the one it gives for the empirical fit parameters $(m_\text{emp},b_\text{emp})$:
\begin{equation}
    \label{eq:quantile}
    Q_{\mathcal{H}_0}(m_\text{emp}, b_\text{emp}) = \iint_{-\infty}^\infty P(m,b) \theta(m_\text{emp}, b_\text{emp}) \diff m \diff b \qquad \theta(m_\text{emp}, b_\text{emp}) = \left\{
    \begin{array}{ll}
    1 & P(m,b) \leq P(m_\text{emp},b_\text{emp}) \\
    0 & \, \textrm{otherwise} \\
    \end{array}
    \right. 
\end{equation}
This quantile represents a stochastically robust measure of the difference between surrogate DRF parameter distribution and the empirical DRF parameters.
It is used to accept or reject the null hypotheses that the surrogate models are based on.
We set the significance threshold for the rejection of null hypotheses at $Q_{\mathcal{H}_0}>0.05$.

\subsubsection{Surrogate model production}
\label{sec:surrogate_production}
In this analysis, surrogate models for four null hypotheses are produced to probe the underlying mechanism of the spreading behavior of BRTs and their relationship with the global flight route network.
A large ensemble of \num{5000} realizations is computed for each surrogate model, to reduce the influence of statistical fluctuations.
We use the canonical naming convention put forward in \cite{gauvin_randomized_2020} to describe the surrogate models $M$ associated with the null hypotheses $\mathcal{H}_0$.
Surrogate models are thus defined by the quantities they conserve with respect to the original empirical data.
To make the surrogate models generally comparable to the empirical data, the number of cities infected in the studied time period, the structure of the network, and the identity of previously infected cities are conserved in all models.
The hierarchy of their null hypotheses, conserving progressively more features of the data, is described here.

\begin{enumerate}
    \item $\mathcal{H}_0^1:\, M(A_{ij}, N_\text{inf})$. 
    \textit{The empirical DRF can be reproduced with a class of models that is only based on the structure of the network.}
    This most basic hypothesis is designed to check if the observed DRF is purely an artifact of the flight network structure.
    To produce the surrogate data set, the identities of infected cities are randomly re-assigned to other cities in the network, and their respective infection times are drawn from a uniform distribution.
    \item $\mathcal{H}_0^2:\, M(A_{ij}, p(T))$. 
    \textit{The empirical DRF can be reproduced with a class of models that is only based on the network structure, and the distribution of infection times.}
    This hypothesis additionally investigates the influence of the infection year distribution, which can be seen to have a strong upward trend in Fig.\,\ref{fig:emp_DRF}D.
    The first step to producing the surrogate model data for this null hypothesis is the randomization of the identity of newly infected cities, analogous to the previous case ($\mathcal{H}_0^1$).
    Here, however, the infection times are drawn from a probability distribution that is derived from the empirical data through kernel density estimation. 
    \item $\mathcal{H}_0^3:\, M(A_{ij}, N_i)$. 
    \textit{The empirical DRF can be reproduced with a class of models that is only based on the network structure, and the position of the infected cities in the network.}
    This hypothesis again builds on ($\mathcal{H}_0^1$), but additionally conserves the identity of newly infected cities.
    It thus tests whether the position of infected cities in the network is sufficient to explain the observed DRF.
    This represents a useful test for homophilic effects: If the null hypothesis cannot be rejected, then it is apparent that the \textit{timing} of the infections is not a relevant factor for the spreading process. 
    Consequently, the probability of becoming infected would not depend on other infections in the network neighborhood. 
    It could instead be dominated by, e.g., the membership in a certain closely-connected clique of cities for which a BRT is especially suitable.
    The surrogate model data is produced by re-drawing the infection year for each city infected in the studied time period from a uniform distribution.
    \item $\mathcal{H}_0^4:\, M(A_{ij}, N_i, p(T))$. 
    \textit{The empirical DRF can be reproduced with a class of models that is only based on the network structure, the network position of the infected cities, and the distribution of infection times.}
    Building on $\mathcal{H}_0^2$ and $\mathcal{H}_0^3$, the overall distribution of infection years is again conserved by this null hypothesis.
    If any artifacts are introduced by the uniform infection time distribution in $\mathcal{H}_0^3$, they should be removed by requiring the infection years to be distributed as they are in the empirical data. 
    This is again achieved by drawing them from a distribution derived from the empirical data through kernel density estimation.
\end{enumerate}

\section{Results}
\label{sec:results}
In this section, the results of the analysis are displayed and interpreted. 
As in Sect.\,\ref{sec:methods}, the empirical dose response function (DRF) is treated first (Sect.\,\ref{sec:results_drf}), followed by the results of the surrogate model study (Sect.\,\ref{sec:results_surrogate}).

\subsection{Empirical Dose Response Function}
\label{sec:results_drf}
\begin{SCfigure}
    \centering
    \includegraphics[width=0.6\linewidth]{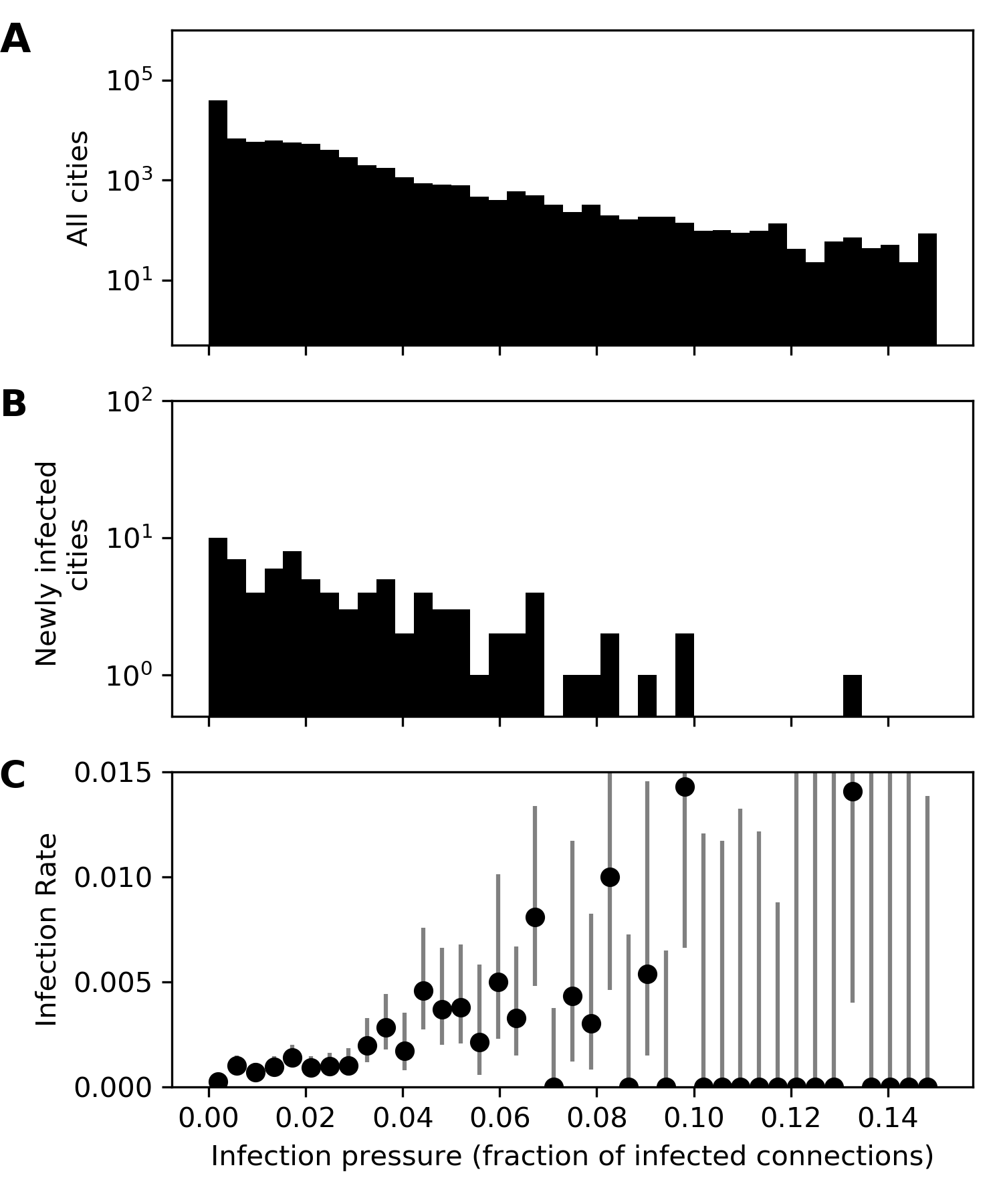}
    \caption{Distributions of \textbf{A} the infection pressures $I_i(t)$ experienced by all cities within the studied time interval, and of \textbf{B} infection pressures experienced by those cities that implemented a Bus Rapid Transit System (BRT) the following year. The infection pressure experienced by a city is defined as the fraction of weighted network connections to cities that have adopted the BRT innovation. In \textbf{C}, the empirical dose response function $p_\text{inf}(I)$ is displayed, obtained by the bin-wise division of \textbf{B} and \textbf{A}. The binomial error of each data point is estimated using Agresti-Coull intervals. A clear upward trend is visible, corresponding to a higher BRT implementation rate for cities whose network neighborhood already featured more BRT implementations.}
    \label{fig:emp_DRF}
\end{SCfigure}

The distribution of infection pressure (dose) values $I$, experienced by any city at any time step, is displayed in Fig.\,\ref{fig:emp_DRF}A. 
The lowest bin, containing cities that had no or very light connections to cities with BRT, is strongly elevated, followed by a brief plateau up to an infection pressure of about 0.03.
Above that value, a roughly exponential decay in frequency from low to high pressures is visible.
``Successful'' infection pressures, that is, infection pressure experienced by those cities that are about to adopt a Bus Rapid Transit System (BRT), are displayed in Fig.\,\ref{fig:emp_DRF}B.
The limited number of BRT adoptions becomes very apparent here; the number of cities newly infected in the studied time period, and thus the number of data points in Fig.\,\ref{fig:emp_DRF}B, is 86.
The downward trend can nevertheless be clearly observed to be flatter than the one in Fig.\,\ref{fig:emp_DRF}A.
Likewise, the lowest bin is not nearly as emphasized.
Dividing the two histograms to obtain the empirical DRF thus yields a graph with a marked upward trend.
The shape of the DRF appears roughly linear, confirming our expectation and justifying the linear fits performed for the surrogate data study (described in Sect.\ref{sec:methods_surrogatedata}).
While the DRF values for many high infection pressures are zero, the very large error bars show that a lack of data introduces significant uncertainty in this regime of larger $I$.

The strong positive correlation between infection pressure and infection rate points to contagion effects at first glance: Cities whose network neighbors have previously adopted BRTs to a large fraction, more frequently also adopt BRTs.
However, this correlation is not necessarily causal.
The correlation could be an artifact of some other process, such as homophilic effects:
If BRTs were especially suited for a certain clique of cities, which happens to be closely connected internally, similar correlations might arise.
There may also be the possibility of other, more basic attributes of the data such as the network structure itself being sufficient in explaining the observed DRF.
To investigate and exclude these confounding effects, surrogate model tests are performed, as discussed in the following section.

\subsection{Surrogate Models}
\label{sec:results_surrogate}
In this section, the results of the surrogate model tests are presented.
For orientation, the short descriptions of the surrogate models are repeated here; they are described in more detail in Sect.\,\ref{sec:surrogate_production}
As mentioned there, the number of cities infected in the studied time period, the structure of the network, and the identity of cities infected before the studied time interval are conserved in all models.

\begin{figure}
    \centering
    \includegraphics[width=0.495\linewidth]{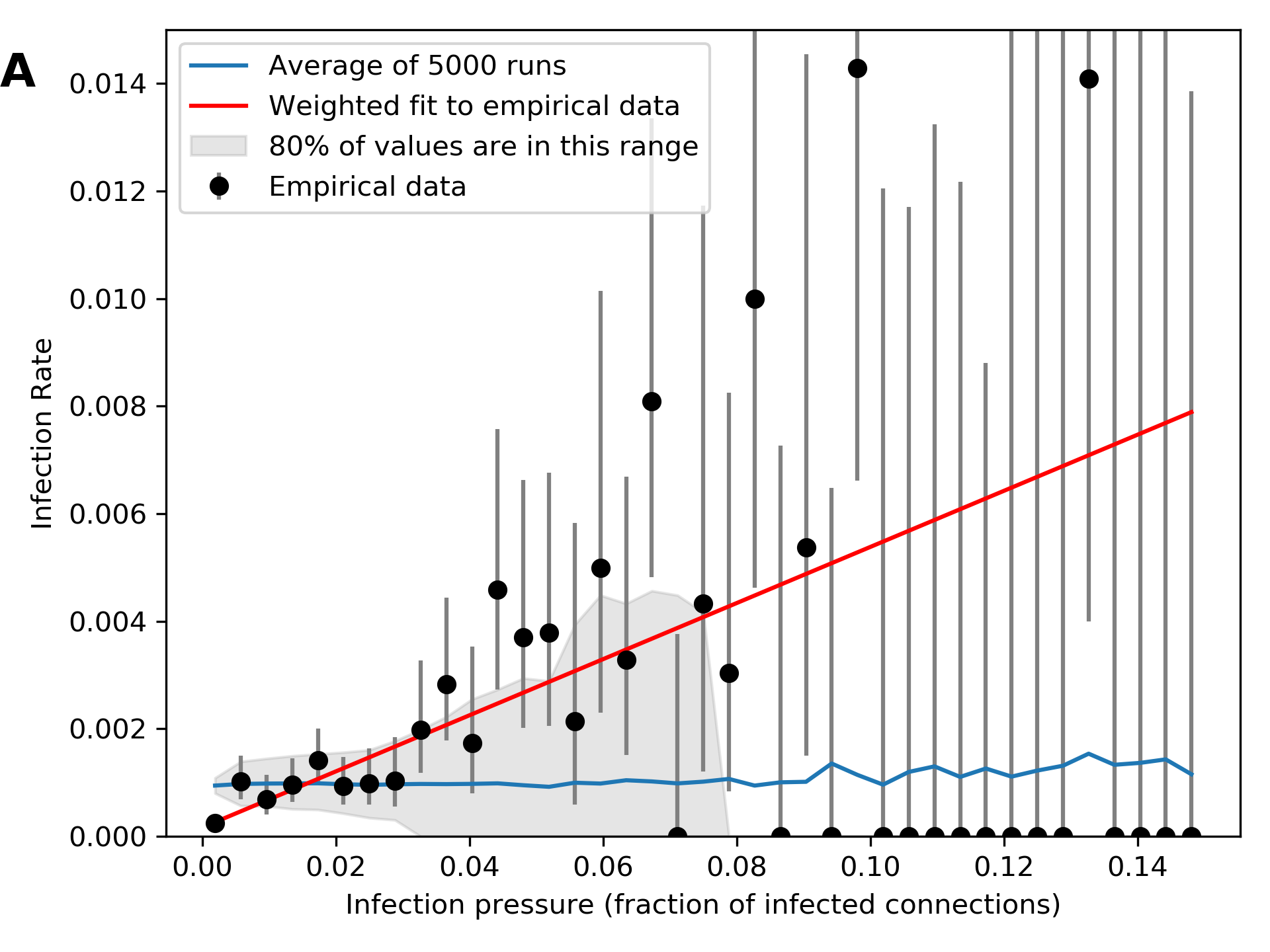}        
    \includegraphics[width=0.495\linewidth]{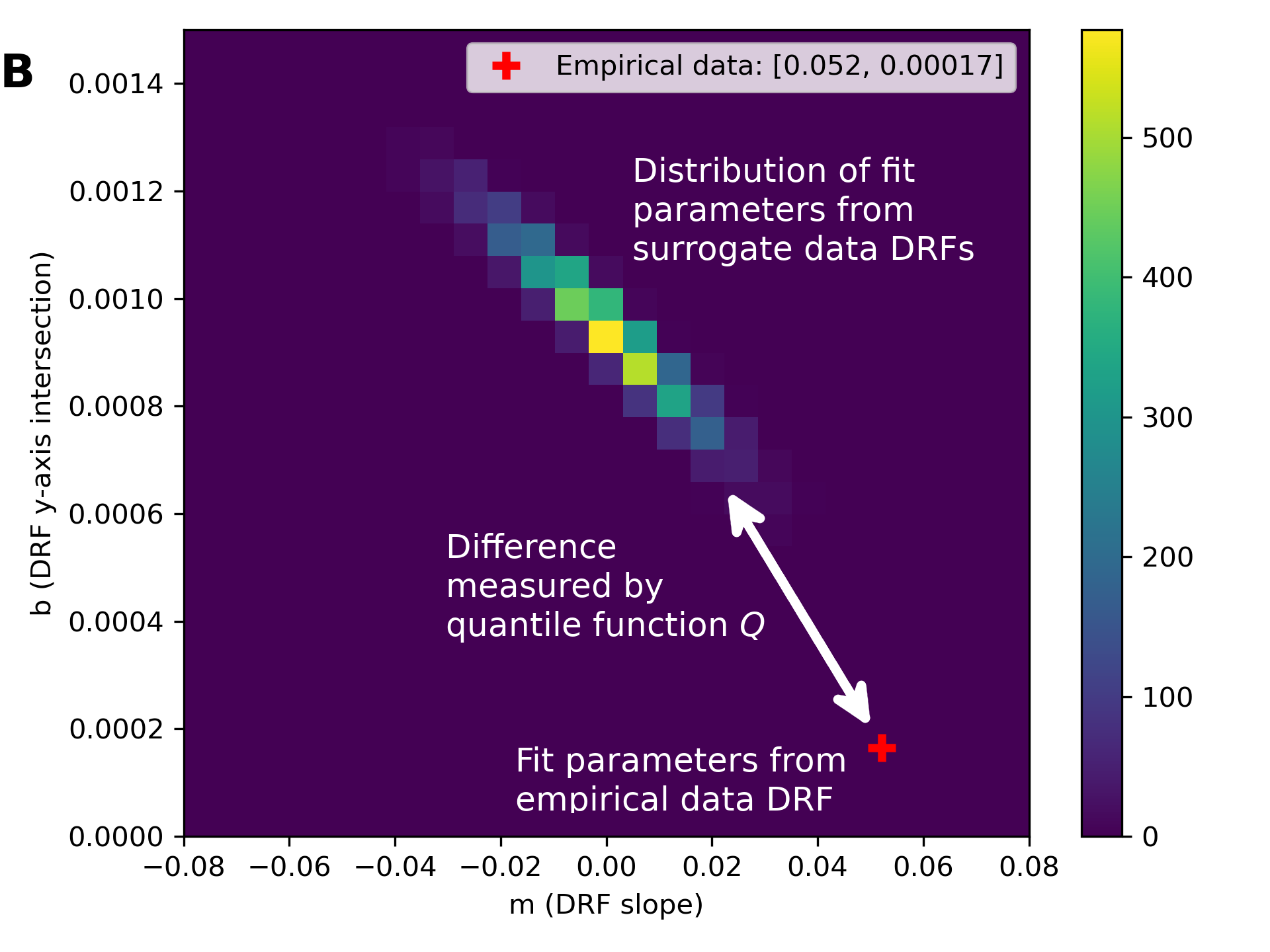}
    \includegraphics[width=0.495\linewidth]{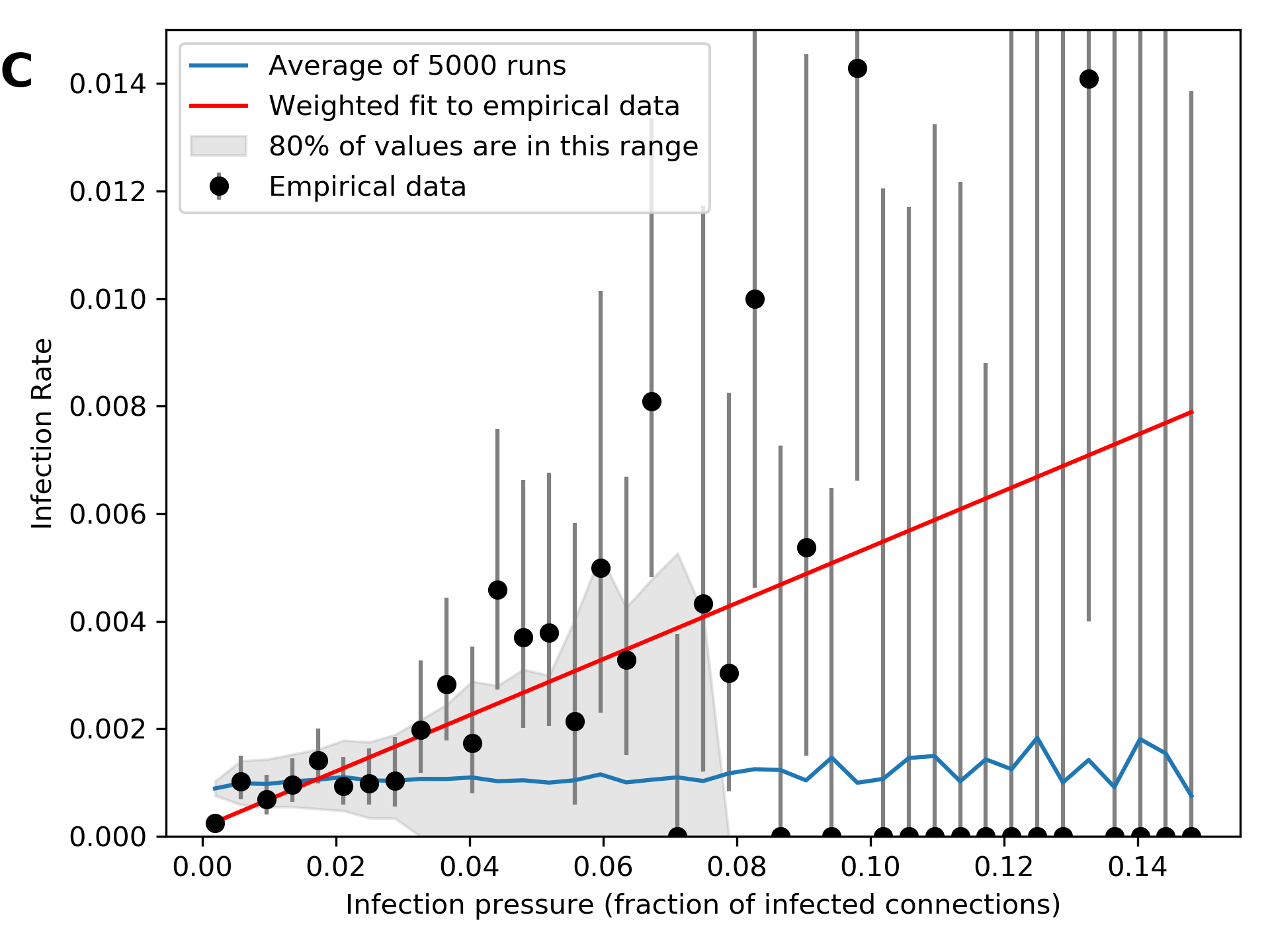}        
    \includegraphics[width=0.495\linewidth]{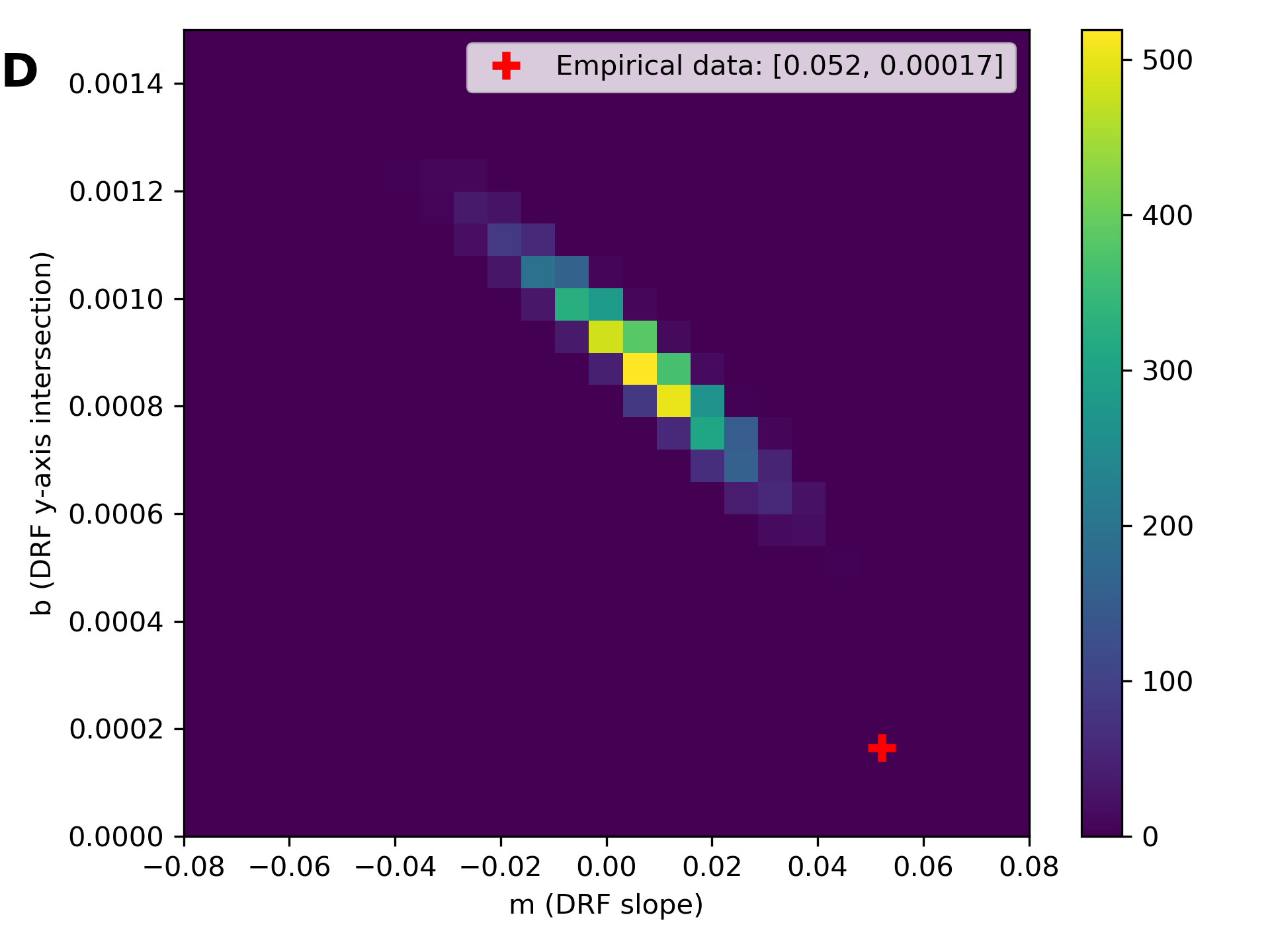}
    \caption{\textbf{A,C} Comparison of the empirical dose response function (DRF, black circles) and the bin-wise average of the DRFs of \num{5000} surrogate model runs (blue), for the surrogate models where the identity of cities implementing BRT (``infected cities'') are randomly reassigned to cities in the network. In \textbf{A} ($\mathcal{H}_0^1$), the infection times (BRT implementation timestamps) of the infected cities are randomized as well, while in \textbf{C}, ($\mathcal{H}_0^2$) the distribution of infection times is conserved.
    The error bars of the empirical DRF are calculated as described in Sect. \ref{sec:errorbars}. 
    Weighted linear least-squares fits are computed for each surrogate model realization; their fit parameters are displayed in \textbf{B} and \textbf{D}. 
    The weighted linear least-squares fit to the empirical data is displayed as a red line in \textbf{A} and \textbf{C}, whose parameters are marked as a red cross in \textbf{B} and \textbf{D}, respectively.
    As opposed to the empirical DRFs, the average surrogate DRFs appear flat throughout most of the infection pressure range (\textbf{A, C}) and the corresponding fit parameter distributions differ strongly from the empirical fit parameters.
    The null hypotheses $\mathcal{H}_0^1$ and $\mathcal{H}_0^2$ can thus be rejected.
    }
    \label{fig:surrogates_randCities}
\end{figure}

\paragraph{First surrogate test. $\mathcal{H}_0^1:\, M(A_{ij}, N_\text{inf})$.} 
\textit{The empirical DRF can be reproduced with a class of models that is only based on the structure of the network, and the number of newly infected cities $N_\text{inf}$.}
This represents the most basic assumption testing whether the observed DRF is merely a product of the structure of the network structure itself, without regard for which or when cities implement a BRT.
This would imply that the spreading of the BRT innovation proceeds completely independent of the network, and thus also completely independent of any other variables correlated with a city's position in the network.
If, instead, a city's network connections are in some way relevant for its BRT adoption probability, the positive correlation of the empirical DRF should not be found in this surrogate model.
As shown in Fig.\,\ref{fig:surrogates_randCities}A, the latter is evidently the case: The average surrogate DRF (in blue) is nearly completely flat over the entire infection pressure range.
The fit parameter comparison in Fig.\,\ref{fig:surrogates_randCities}B thus only confirms the obvious.
The empirical DRF's fit parameters are strongly separated, with $Q_{\mathcal{H}_0^1}\approx0$.
The null hypothesis is thus rejected.

\paragraph{Second surrogate test. $\mathcal{H}_0^2:\, M(A_{ij}, p(T))$.}
\textit{The empirical DRF can be reproduced with a class of models that is only based on the network structure, and the distribution of infection times $p(T)$.}
The second null hypothesis builds on the first one, and additionally preserves the yearly BRT adoption rate; that is, the distribution of the years in which new BRTs are implemented.
In view of the strong rejection of $\mathcal{H}_0^1$, we expect a similar result for this surrogate model, since the potential effect of the variable infection rate appears negligible compared to the randomization of which cities become infected.
The result of this test is displayed in Fig.\,\ref{fig:surrogates_randCities}C and D.
As expected, the average surrogate model DRF remains flat, and the distribution of surrogate DRF fit parameters is strongly separated from the empirical DRF fit parameters.
A difference to $\mathcal{H}_0^1$ is visible, however: 
the fit parameter distribution (Fig.\,\ref{fig:surrogates_randCities}D) is shifted towards positive slopes, whereas it was centered around $m=0$ for $\mathcal{H}_0^1$ (Fig.\,\ref{fig:surrogates_randCities}B).
Combined with the correspondingly lower values for $b$, this reduces the separation between the distribution and the empirical for parameters.
However, the value of the quantile function remains $Q_{\mathcal{H}_0^2}\approx0$, and $\mathcal{H}_0^2$ is thus rejected.
The time distribution of the BRT implementations appears to hold significance for the DRF, but is not nearly sufficient to explain the empirically observed correlation.

\paragraph{Third surrogate test. $\mathcal{H}_0^3:\, M(A_{ij}, N_i)$.}
\textit{The empirical DRF can be reproduced with a class of models that is only based on the network structure, and the position of the infected cities in the network.}
This hypothesis again builds on \Hone, this time by conserving the the identity of the infected nodes in the network, meaning which of the cities implement BRT.
The resulting surrogate model is thus much more strongly constrained to resemble the empirical data than for the previous two hypotheses. 
The times at which these infections occur are randomly drawn from a uniform distribution.
Using this null hypothesis, we can probe the system for homophilic effects.
If, for example, there is a closely connected clique of cities that BRT is especially suited for, or that share a BRT-favorable political environment, then the infected cities' position in the network would be expected to be sufficient to reproduce the observed DRF.
In particular, the specific times of infection / BRT adoption would be irrelevant, as the null hypothesis posits.

\begin{figure}
    \centering
    \includegraphics[width=0.495\linewidth]{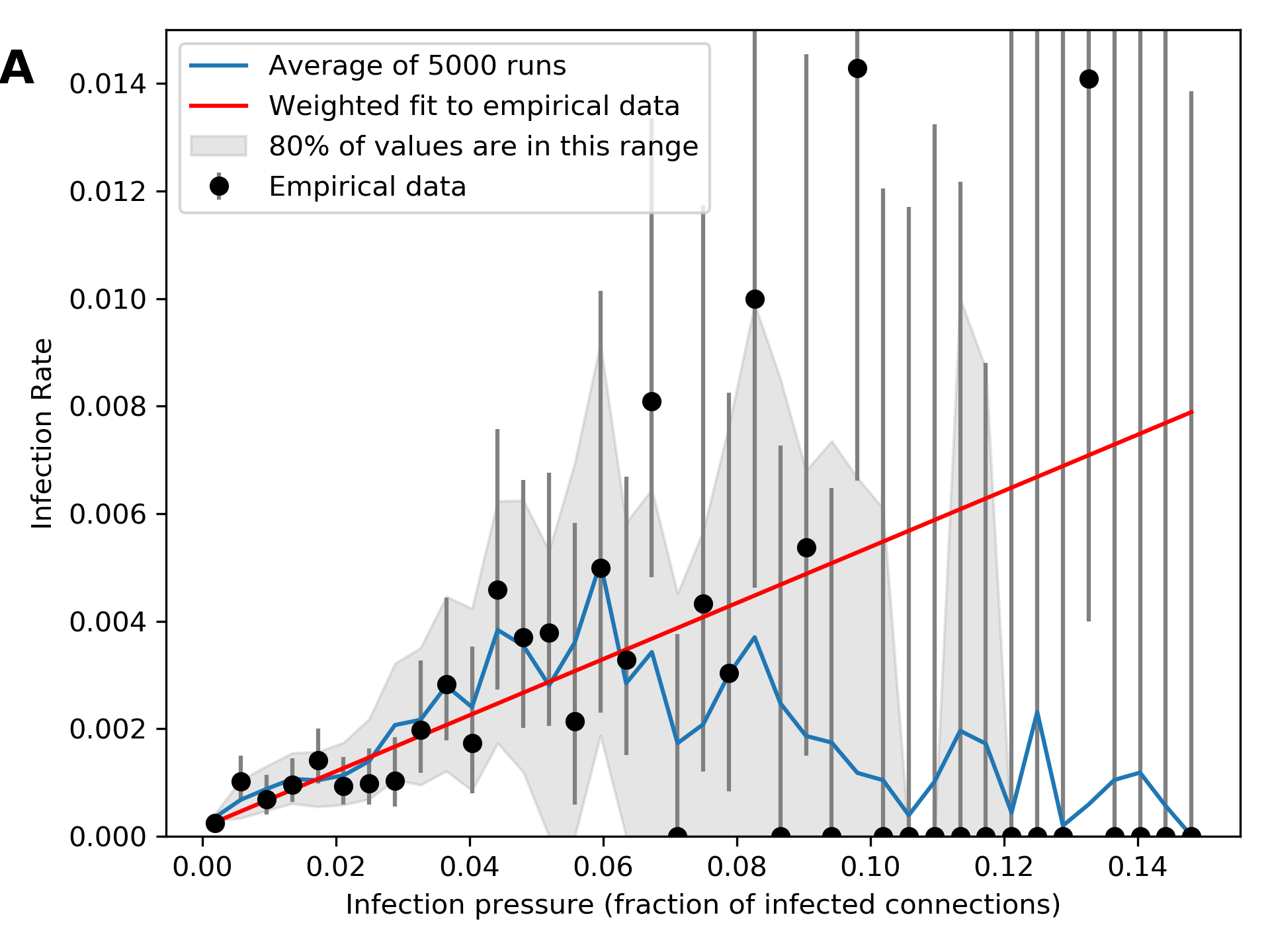}
    \includegraphics[width=0.495\linewidth]{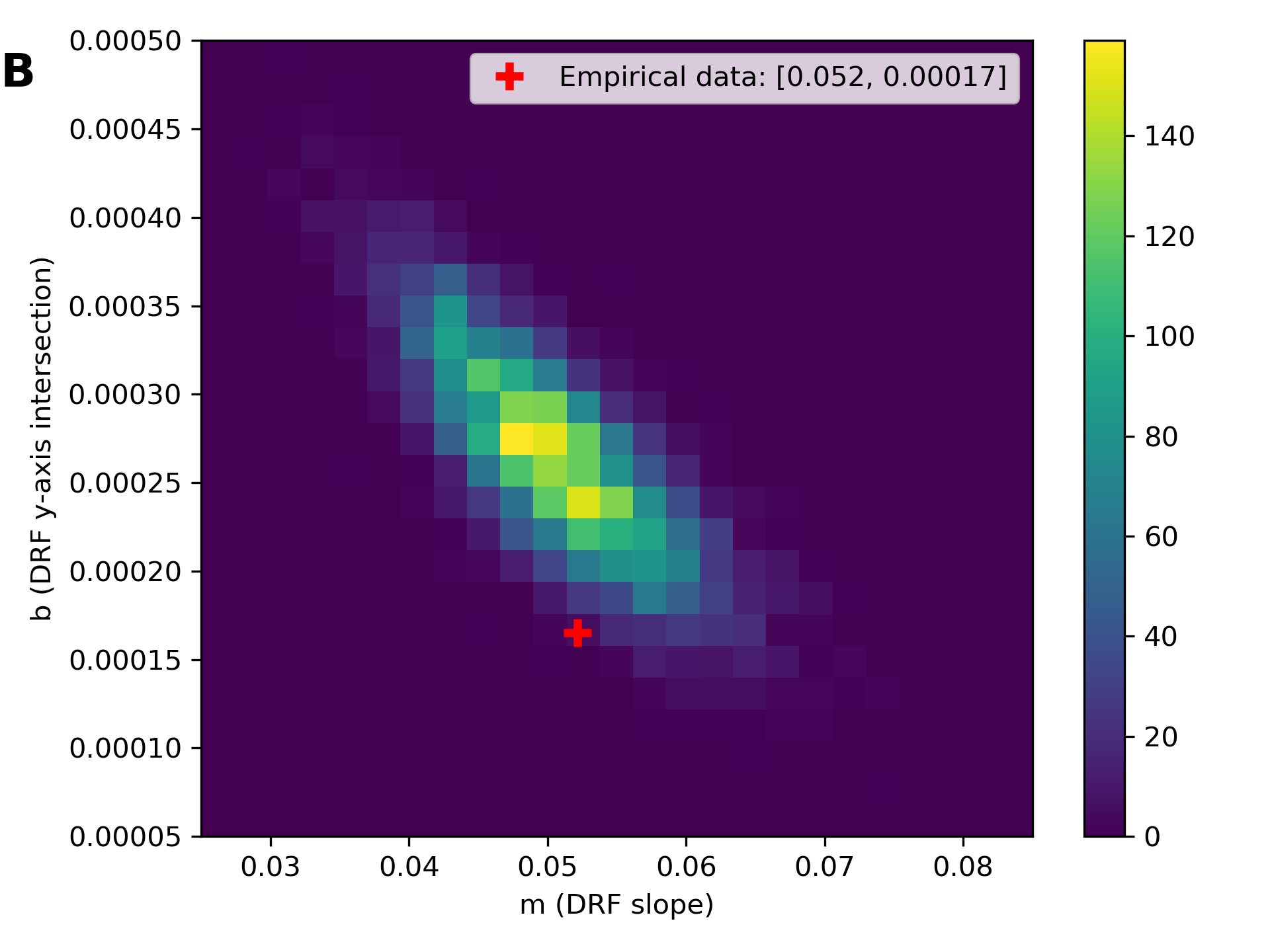}
    \includegraphics[width=0.495\linewidth]{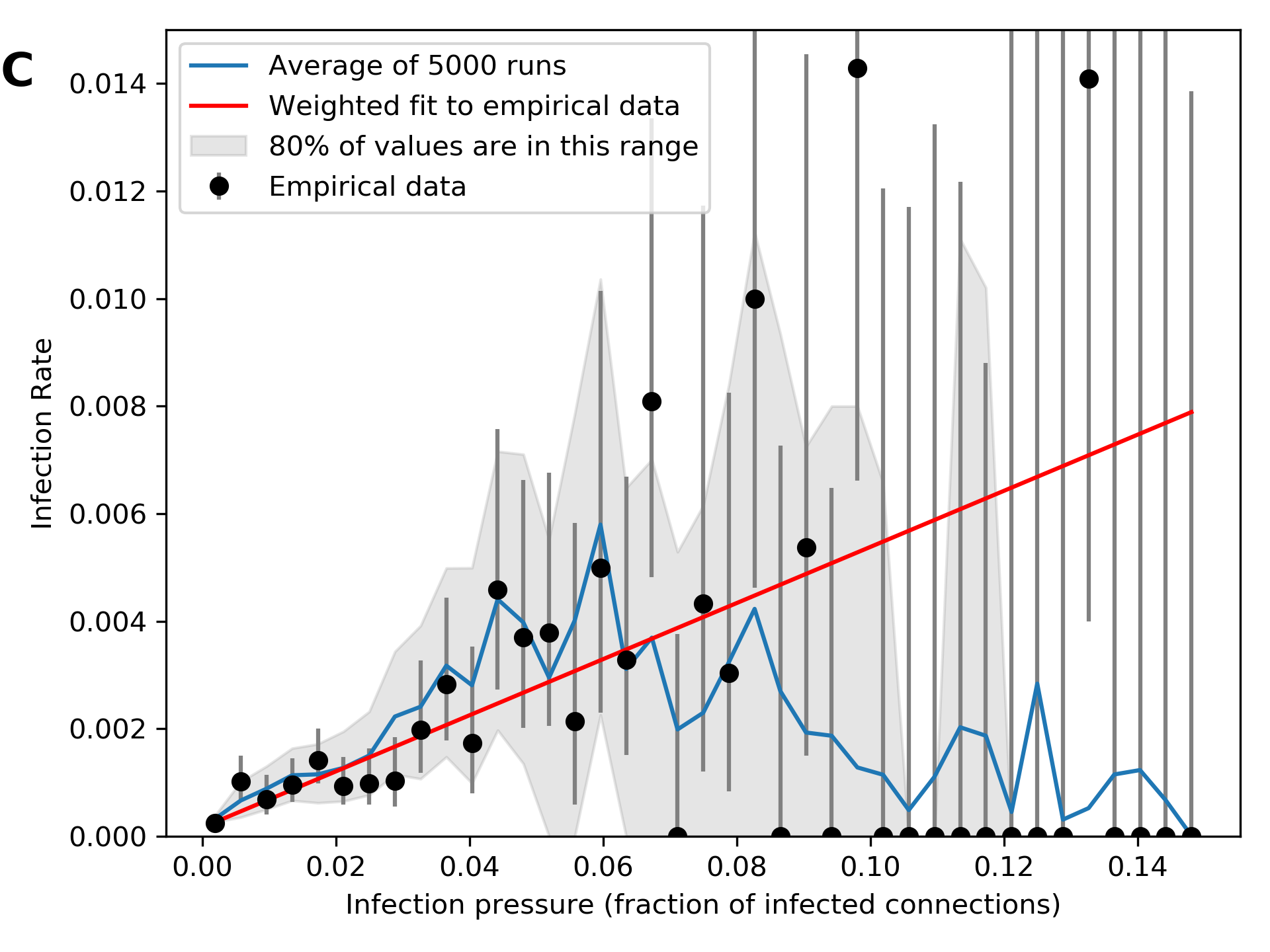}        
    \includegraphics[width=0.495\linewidth]{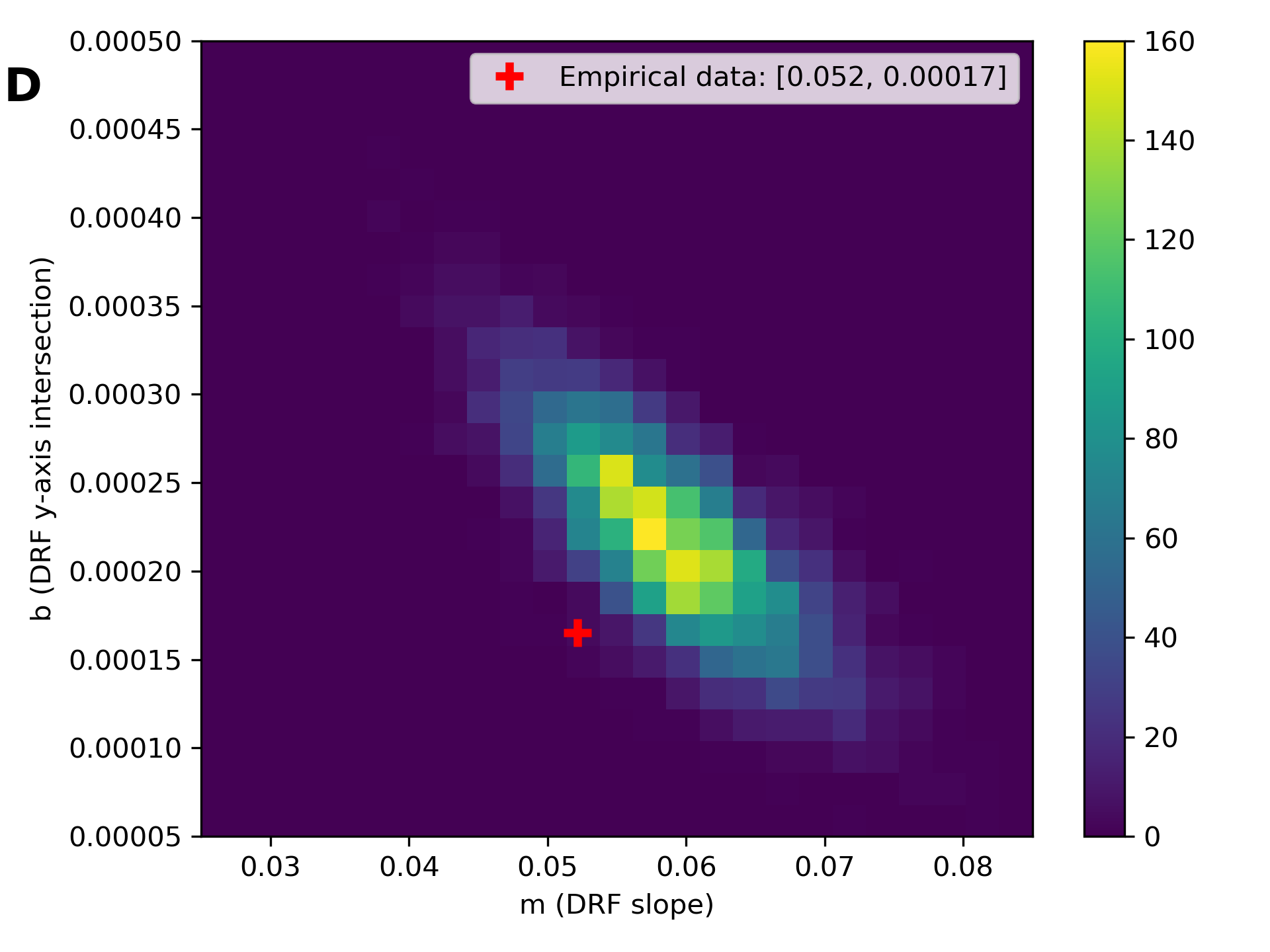}
    \caption{\textbf{A,C} Comparison of the empirical dose response function (DRF, black circles) and the bin-wise average of the DRFs of \num{5000} surrogate model runs (blue), for the surrogate models where the identity of cities implementing BRT (``infected cities'') are conserved. 
    In \textbf{A} ($\mathcal{H}_0^3$), the infection times of the infected cities are randomly drawn from a uniform distribution covering the investigated time interval.
    In \textbf{C}, ($\mathcal{H}_0^4$) the infection times are drawn from a distribution derived from a kernel density estimation of the distribution found in empirical data. 
    The error bars of the empirical DRF are calculated as described in Sect. \ref{sec:errorbars}. 
    Weighted linear least-squares fits are computed for each surrogate model realization; their fit parameters are displayed in \textbf{B} ($\mathcal{H}_0^3$) and \textbf{D} ($\mathcal{H}_0^4$). 
    The fit to the empirical data is displayed as a red line in \textbf{A} and \textbf{C}, whose parameters are marked as a red cross in \textbf{B} and \textbf{D}.
    In both \textbf{A} and \textbf{C}, the average surrogate DRFs appear to match the empirical DRF.
    However, individual data points are correlated within each surrogate model realization's DRF.
    The fit parameter comparison in \textbf{B} and \textbf{D} takes this into account, revealing a significant difference between the distribution of surrogate model fit parameters and those of the empirical DRF.
    The null hypotheses $\mathcal{H}_0^3$ and $\mathcal{H}_0^4$ can thus be rejected as well.
    }
    \label{fig:surrogates_constCities}
\end{figure}

The comparison of the empirical DRF with the ensemble of surrogate model DRFs is displayed in Fig.\,\ref{fig:surrogates_constCities}A and B.
Here, the importance of not relying on the bin-wise average of the surrogate DRF ensemble becomes apparent: while the difference between empirical and surrogate data is not readily apparent in Fig.\,\ref{fig:surrogates_constCities}A, it can more clearly be discerned in B.
While the value of the quantile function $Q_{\mathcal{H}_0^3}=0.0271$ comes closer to the chosen significance threshold than in the previous two tests, the null hypothesis is nonetheless rejected.
Evidently, this surrogate model can reproduce the empirical DRF much better than the two previous ones.
Which nodes become infected is thus apparently correlated with their network position, and homophilic effects may be at play.
However, they do not appear sufficient on their own to explain the observed empirical DRF.

We observe that the \textit{individual} values of $m_\text{emp}$, and to a lesser extent $b_\text{emp}$, are not uncommon in the fit parameter distribution of the surrogate DRFs.
Only in the two-dimensional graph does it become apparent that their \textit{combination} $(m_\text{emp}, b_\text{emp})$ is very rarely observed.
It appears that the empirical DRF's data point at $I\approx0$, being very close to $p_\text{inf}(I) = 0$ and having very small error bars (Fig.\,\ref{fig:surrogates_constCities}A), more strongly constrains $b_\text{emp}$ to small values than is the case for the surrogate model data.
We interpret this as the observation that, in the absence of an infected, or ``seed'' city in a city's neighborhood, the infection rate is inhibited.
Thus, in the empirical data, cities are unlikely to adopt BRT if \textit{none} of their neighbors have done so yet, suggesting a contagion process to be possibly at play.

The analysis of \Htwo demonstrated that the non-constant rate of infections may play a role for the underlying process. 
Building on \Hthree, but conserving this distribution as well, thus remains as the final test that may distinguish between homophily and potential contagion effects.
This most restrictive surrogate analysis performed in this study is described in the following paragraph.

\paragraph{Fourth surrogate test. $\mathcal{H}_0^4:\, M(A_{ij}, N_i,  p(T))$.} 
\textit{The empirical DRF can be reproduced by a class of models that is only based on the network structure, the network position of the infected cities, and the distribution of infection times.}
Combining null hypotheses \Htwo and \Hthree, this hypothesis is designed as the most stringent test for homophily.
The infection timestamps of newly infected cities are, again, sampled from a probability distribution derived from empirical data via kernel density estimation. 

The results of this test are displayed in Fig.\,\ref{fig:surrogates_constCities}C and D.
Similar to \Hthree, no significant difference between the empirical and average surrogate DRFs can be observed (Fig.\,\ref{fig:surrogates_constCities}C).
However, the distribution of surrogate DRF fit parameters is clearly removed from the empirical DRF's fit parameter. 
The additional conservation of the infection time distribution has shifted the distribution towards higher $m$ and lower $b$, similar to the difference between \Hone and \Htwo.
With a quantile function value of $Q_{\mathcal{H}_0^4}=0.0143$, null hypothesis \Hfour is actually somewhat more strongly rejected than \Hthree.

Similarly to \Hthree, the individual values for $m_\text{emp}$ and $b_\text{emp}$ are common in the distribution of surrogate DRF fit parameters, and steeper DRF slopes $m>m_\text{emp}$ are actually more likely to occur.
However, the combined analysis of $(m_\text{emp},b_\text{emp})$ again shows the significant difference between the empirical and surrogate data.
This demonstrates the strength of the method in distinguishing differences between empirical and surrogate data, beyond what would have been visually perceivable in Fig.\,\ref{fig:surrogates_constCities}C.

Both contagion and homophilic mechanisms are expected to create a strong correlation between the nodes' network position and their probability to become infected.
However, only in the case of contagion dynamics does the relative timing of infection events matter:
For a kind of ``infection wave'' traveling from node to node across the network, the order of infections is not random.
Thus, with the rejection of both \Hthree and \Hfour, we find homophilic effects to be an insufficient explanation of the data.
We interpret this as evidence pointing towards underlying contagion mechanisms controlling the spreading behavior of BRT innovations between cities, where the linkages provided by the global flight route network appear to be an adequate proxy.

\begin{SCtable}
    \centering
    \begin{tabular}{cl|l}
         Null Hypothesis & Surrogate Model & $Q_{\mathcal{H}_0^i}$ \\
         \midrule
         \Hone  &   $M(A_{ij}, N_\text{inf})$   & \num{0.0}     \\
         \Htwo  &   $M(A_{ij}, p(T))$           & \num{0.0}     \\
         \Hthree&   $M(A_{ij}, N_i)$            & \num{0.0271}   \\
         \Hfour &   $M(A_{ij}, N_i, p(T))$      & \num{0.0143}  \\
    \end{tabular}
    \caption{Values of the quantile function $Q(m_\text{emp}, b_\text{emp})$, as defined in Eq.\,\ref{eq:quantile} for the four investigated null hypotheses. All four are rejected, though much more narrowly for \Hthree and \Hfour. For details on the kernel density estimations which form the basis for the calculation of Q, see App.\,\ref{app:smoothed_fit_params}.}
    \label{tab:Q_values}
\end{SCtable}

\section{Discussion and Conclusion}
\label{sec:discussion}
Cities may learn from each other which policies and technologies to adopt.
As part of the human response to anthropogenic environmental changes, this represents an Earth System process that has attracted little research attention at the macroscopic scale so far.
In this contribution, we proposed a method for investigating the spreading of urban innovations related to sustainability on the global network of cities.
The method is made up of two steps:
First, we estimate dose-response functions (DRFs) from empirical data on the network proxy and spreading process.
Second, we perform hypothesis tests on surrogate data models generated from the empirical data, to probe and exclude specific effects that may confound the detection of contagion effects.

This method was demonstrated on a pair of example data sets:
We correlate the spreading of Bus Rapid Transit Systems, a public transport innovation, with the global network of flight routes as a proxy for inter-city learning connections.
We find significant evidence towards contagion processes in this data, which cannot be sufficiently explained by homophilic effects such as a shared environment or clustering.
Cities whose neighborhood is free of BRT implementations appear especially unlikely to adopt a BRT themselves.
This underlines the fact that cities often base their policy decisions on each others' experiences.

Our results indicate that it is possible to find proxies for the global city learning network.
This forms the basis for future work, for example on comparing which inter-city linkages offer the largest predictive power for the spreading of innovations, and what kind of (complex) contagion processes are at work.
The investigation of these processes has the potential to improve our understanding of the human components of the Earth System. 
Identifying ``gaps'' in this network, such as cluster boundaries that innovations rarely cross, may even open the doors for targeted interventions in the future.
Such actions could ensure that vital innovations are adopted quickly around the globe, and reduce the time lag between the inception of a good idea and its environmental impact on a global scale.

Our analysis has a range of limitations, which we would like to address here.
Firstly, those relating to the datasets used to illustrate the method.
We base the network component of our analysis, on a publicly available dataset of scheduled flight routes, which provides only a static snapshot.
Temporally resolved data, especially on actual origin-to-destination passenger flows instead of scheduled airport-to-airport routes, could be expected to contain a clearer signal.
On the matter of the spreading innovation, the low ($\mathcal{O}(100)$) number of total BRT adoptions likely significantly reduces our ability to discern actual trends from noise and fluctuations in the data.
On the methodological side, our analysis is limited by the proof-by-contradiction nature of our method: contagion processes may only be indirectly inferred.
Furthermore, DRFs are a highly aggregate statistical measure, which may not be specific enough for accurately characterizing subtle spreading processes under the data limitations.
A systematic analysis of the statistical power of the method under varying conditions of noise, system size, different underlying spreading processes, and other parameters would be desirable to more accurately interpret the results.
Such an analysis, which could be achieved using large ensembles of simulations, lies beyond the scope of this contribution.

While our method can exclude specific correlations and causal connections in the data, it cannot provide a detailed process-based understanding of of urban innovation transmission.
In future work, hypotheses of concrete mechanisms should be tested, for example using combinations of Monte Carlo and maximum likelihood methods.
A similarly promising approach would be to use higher-order statistics, such as multi-node correlations for identifying longer contagion chains and network motifs related to spreading dynamics.

The applications of our method is limited by data availability.
Comprehensive databases on inter-city linkages are rare, especially on the global scale.
While abundant data often exists on the national and inter-national level, homogeneous city-scale data that covers more than just the largest ``world cities'' \cite{taylor_specification_2001,taylor_measurement_2002} is much harder to find.
If cities' role as active agents in the Anthropocene is to be taken seriously, then the collection of high-quality data on this scale is paramount.
The comparative analysis of multiple spreading processes correlated with a number of different inter-city network proxies, promises valuable insights on the real-world mechanisms underlying the transfer of urban innovations.

Our proposed method is generic, and can be applied to probe for contagion effects in any combination of city network and spreading innovation.
It can also easily be generalized to temporally dynamic networks.
Furthermore, it may be useful for the analysis of other complex systems wherever contagion processes are hypothesized in low-rate spreading phenomena on densely connected networks.
Potential applications include the spreading of opinions, social norms and behaviors among individuals and more abstract agents, wherever such data sets are available.
This highlights a potential connection to research on social tipping processes \cite{otto_social_2020}, which are one of the most promising pathways for mitigating dangerous anthropogenic global warming and the long-term crossing of other planetary boundaries.

\section*{Acknowledgements}
The authors would like to thank Prof. Xuemei Bai for discussions instrumental for the development of this project, and Prof. Ricarda Winkelmann, Dr. Nico Wunderling and Dr. Jobst Heitzig for further helpful discussions.
NHK is grateful for financial support from the Geo.X Young Academy.
JFD is thankful for financial support by the Leibniz Association (project DominoES) and by the European Research Council project Earth Resilience in the Anthropocene (743080 ERA).
NHK, PR and JFD acknowledge the Cooperation and Collective Cognition Network, funded by the Humboldt \& Princeton university strategic partnership, for paving the path leading to this research.
PR has been funded by the DFG (German Research Foundation) - RO 4766/2-1.

The data on Bus Rapid Transit Systems was obtained from BRTData \cite{brtdata} under a \textit{Creative Commons BY-NC-ND 3.0} license \cite{brtdata_license}.
The data on scheduled flight routes was obtained from OpenFlights \cite{openflights}, made available under Open Database \cite{openflights_opendatabase} and Database Contents Licenses \cite{openflights_databasecontents}.
City data sourced from wikidata \cite{wikidata} is in the public domain \cite{wikidata_license}.

\section*{Author contributions}

NHK, PR and JFD conceived this study. 
NHK prepared the data, implemented the data analysis methodology, and visualized the
results, with supervision by JFD and PR. 
All authors interpreted and discussed the results. 
NHK wrote the manuscript, with support from PR and JFD. 
All the authors give their final approval of the article version to be published.

\section*{Code and data availability}

The data analysis scripts and (to the extent that licenses allow) curated data sets used in this study will be made available as open source code upon publication of this article.

\bibliographystyle{naturemag}
\bibliography{bib} 


\newpage
\appendix

\section{Quantile function calculation for fit parameter distributions}
\label{app:smoothed_fit_params}
\begin{figure}
    \centering
    \includegraphics[width=0.495\linewidth]{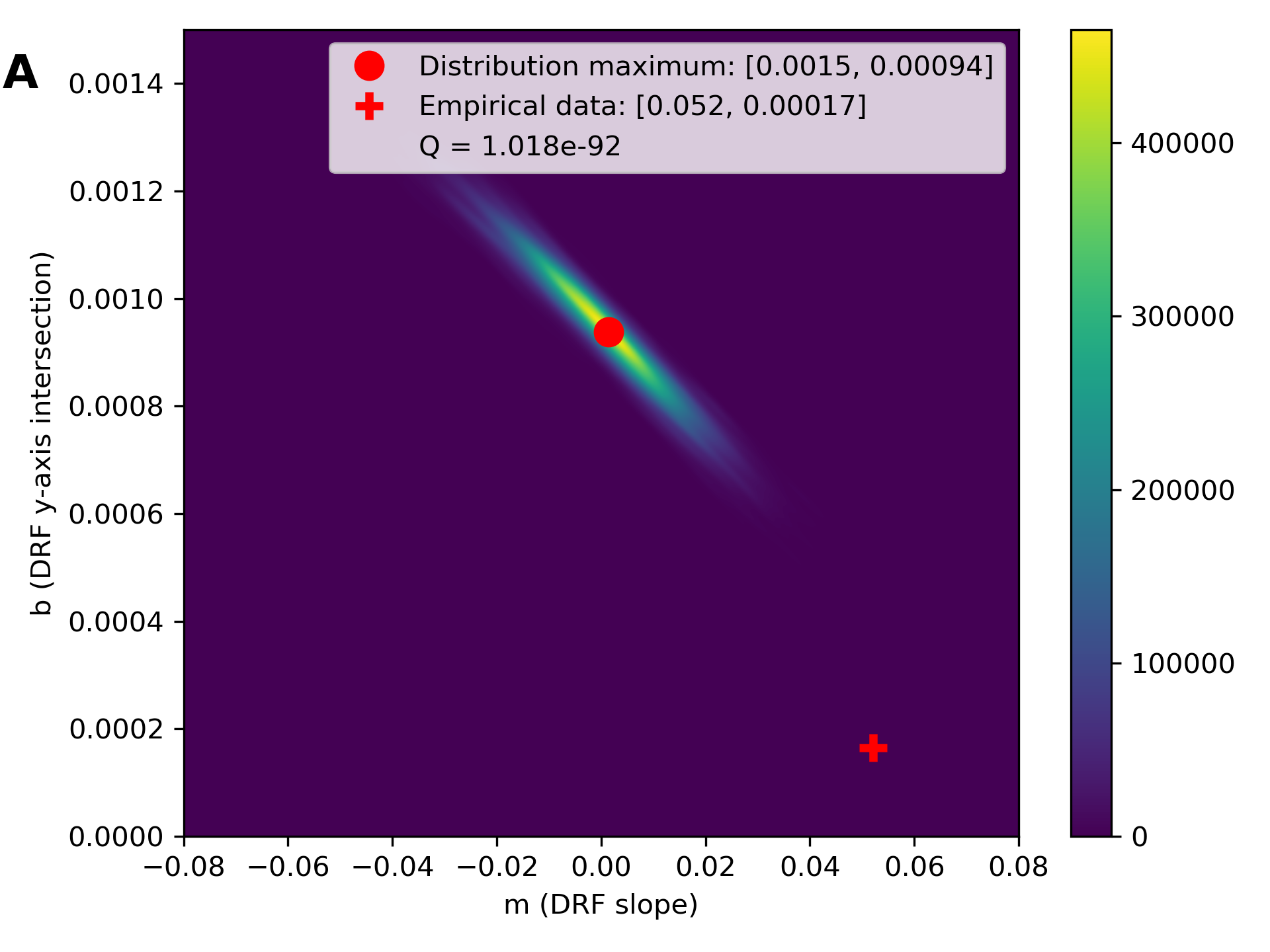}
    \includegraphics[width=0.495\linewidth]{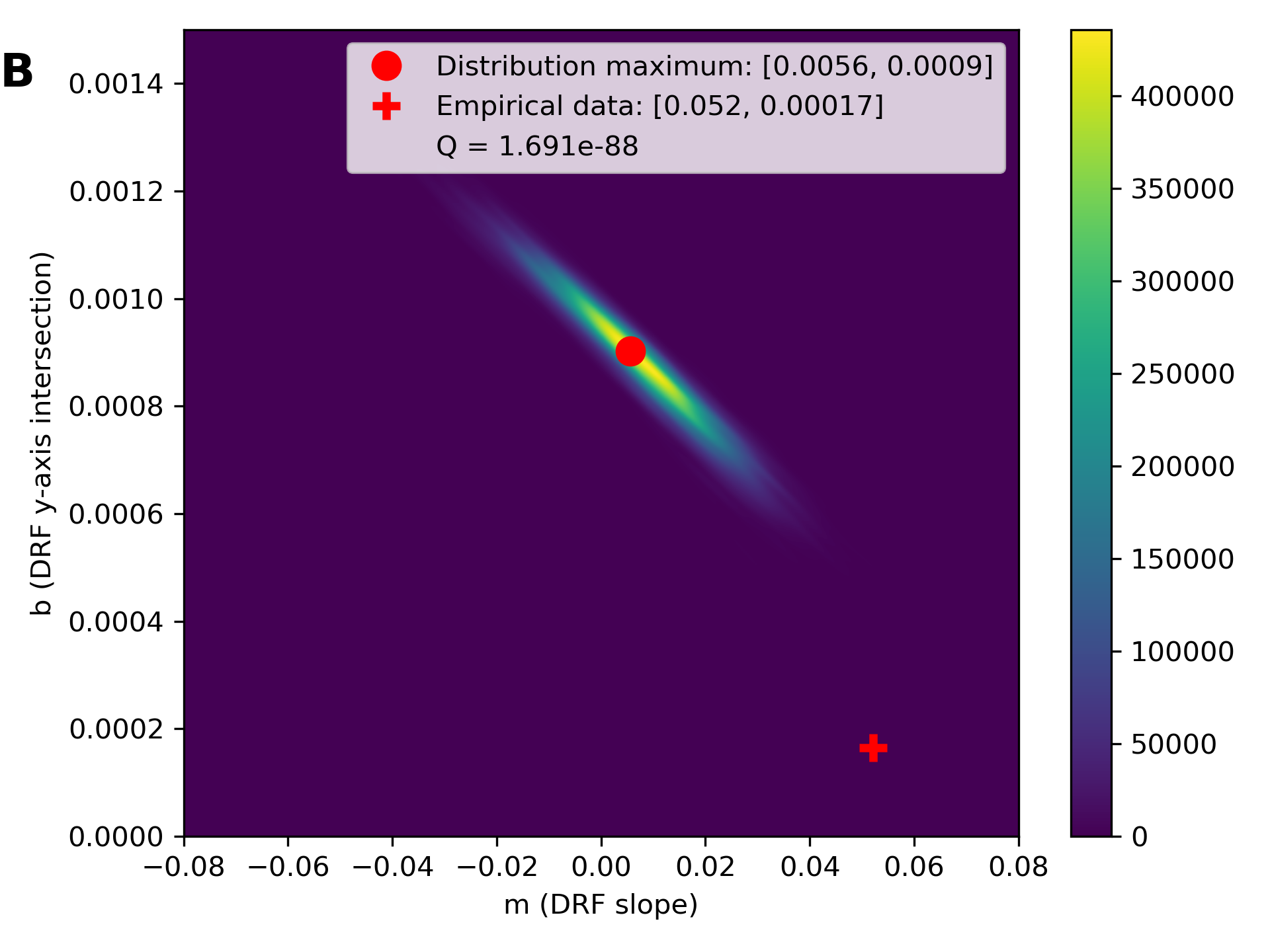} 
    \includegraphics[width=0.495\linewidth]{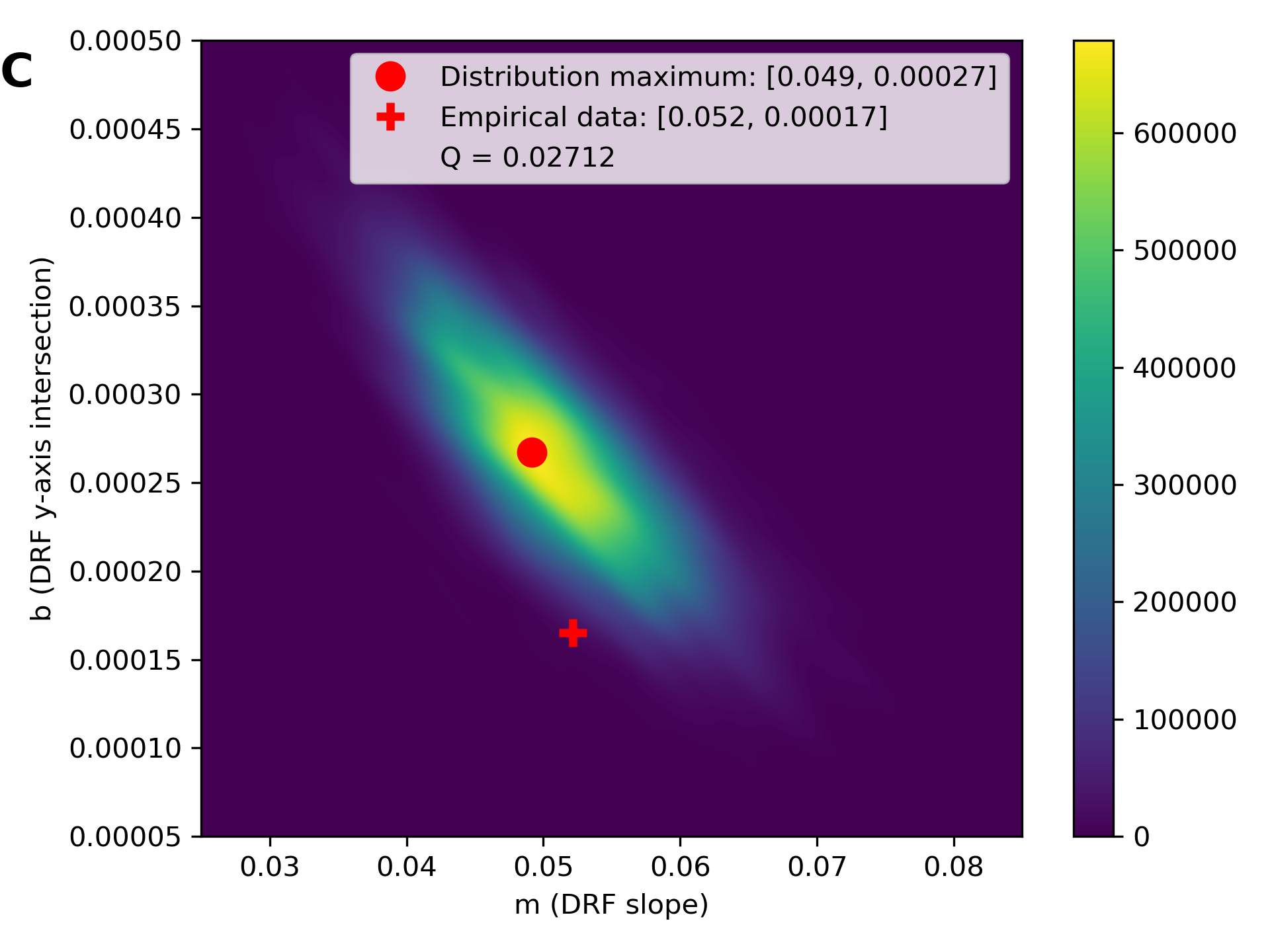}
    \includegraphics[width=0.495\linewidth]{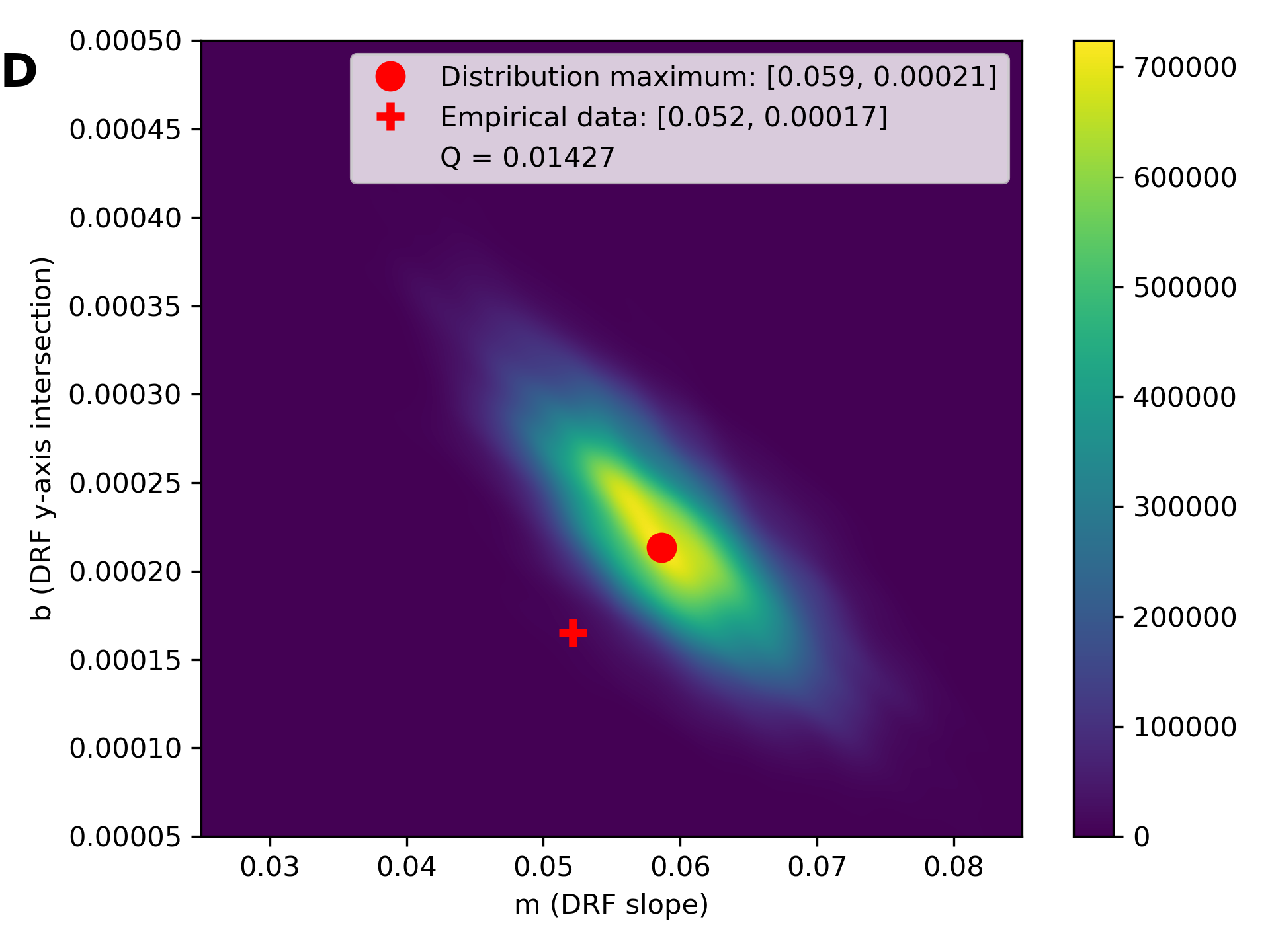}
    \caption{
    Parameter probability distributions of the linear fits on every surrogate data realization. The probability distributions are determined via kernel density estimation, and correspond to the surrogate ensembles for \textbf{A} \Hone, \textbf{B} \Htwo, \textbf{C} \Hthree and \textbf{D} \Hfour.}
    \label{fig:smoothed_fit_params}
\end{figure}

The underlying probability distributions from the fit parameter distributions displayed in Figs.\,\ref{fig:surrogates_randCities} and \ref{fig:surrogates_constCities}, are estimated using kernel density estimations.
Gaussian kernels were used, and their bandwidth were determined using Silverman's rule \cite{silverman_density_1986}.
In Fig.\,\ref{fig:smoothed_fit_params}, the estimated probability distributions are displayed.
The fit parameters of the empirical DRF is shown as a red cross.
The value of the quantile function $Q$, measuring the likelihood of the empirical data point being produced by the distribution, is noted in the figures' legends.

\section{City connection weight calculation}
\label{app:city_distances}
The connection weights of the cities, interpretable as abstract proximity measures, are derived from the flight route database \cite{openflights} and the geographic distance between airports and city centers \cite{openflights, wikidata}.
The network is constructed as follows:
\begin{enumerate}
    \item All cities with a population above \num{60000} are considered as nodes in the network.
    \item All airports within \SI{60}{\kilo\meter} of a city's center are considered to belong to this city. Cities sharing airports and vice versa are possible.
    \item Two cities are connected to each other if one can be reached from the other with $\leq2$ flights. The weight $w_{\text{A,B}}$ of this connection functionally depends on the multiplicity of the flight routes $r_{\text{A$\rightarrow$B}}$ and $r_{\text{B$\rightarrow$A}}$ that directly connect A and B (if any), as well as the number and multiplicity of viable 2-flight paths between the cities. The functional dependence is defined here:
    \begin{align}
        \label{eq:city_proximity}
        w_{\text{A,B}} &= f_{\text{A$\rightarrow$B}} + f_{\text{B$\rightarrow$A}}\\
        f_{\text{A}\rightarrow\text{B}} &= r_{\text{A}\rightarrow\text{B}} + \sum_{k=0}^{N_A} \frac{r_{{\text{A}\rightarrow k}} r_{k{\rightarrow\text{B}}}}{\sum\limits_{l=0}^{N_k} r_{{k\rightarrow l}}} \qquad
        \end{align}
    Here, $\sum_{k=0}^{N_A}$ refers to the sum over all $N_\text{x}$ cities that can be reached with a single flight from city $x$.
    Thus, the contribution of any 2-flight path is the A $\rightarrow$ B $\rightarrow$ C is the product of the route multiplicities $r_{\text{A$\rightarrow$B}}$ and $r_{\text{B$\rightarrow$C}}$, divided y the out-degree of node B.
\end{enumerate}

Potential connections using more than two flights are not considered, as this would significantly increase the computation time and network density, while not adding significantly weighted connections. 
Note that the scheduled flights themselves are directed; A $\rightarrow$ B $\rightarrow$ C represents a valid connection between A and B, while A $\rightarrow$ B $\leftarrow$ C does not.
However, the resulting city connection weight $w_{\text{A,B}}=w_{\text{B,A}}$ is undirected, or symmetric, by construction.

\end{document}